\documentclass[cits]{PoS}

\usepackage{amsmath}		
\usepackage{amssymb}		
\usepackage{amsthm}		

\usepackage{multirow}		
\usepackage{array}		

\usepackage{mathrsfs}		
\usepackage{bm}			
\usepackage{cancel}		

\usepackage{wrapfig}

\newcommand{\resizedgraphics}[2]{
\resizebox{#1\textwidth}{!}{\includegraphics{#2}}
}

\graphicspath{{./figures/}{../../../../../Physica/Isospin_chemical_potential/figures/}%
{../../../../../Physica/Isospin_chemical_potential/figures/Contours_new/}}%

\usepackage{url}			
\usepackage{makeidx}		

\newcommand{\Eq}[1]{Eq.\,\eqref{#1}}
\newcommand{\Eqs}[1]{Eqs.\,\eqref{#1}}
\newcommand{\Fig}[1]{Fig.~\ref{#1}}

\renewcommand{\Re}{\operatorname{\mathfrak{Re}}}
\renewcommand{\Im}{\operatorname{\mathfrak{Im}}}

\newcommand{\abs}[1]{\left|#1\right|}
\newcommand{\ave}[1]{\left\langle#1\right\rangle}
\newcommand{\Ave}[1]{\left\langle\!\!\left\langle#1\right\rangle\!\!\right\rangle}
\newcommand{\AVE}[1]{\left\langle\!\!\!\!\left\langle#1\right\rangle\!\!\!\!\right\rangle}
\newcommand{\RnB}[1]{\left(#1\right)}

\newcommand{\Nf}{N_{\textrm{f}}}			
\newcommand{\Nsite}{N_{\textrm{site}}}	

\newcolumntype{G}{>{$ \displaystyle \raisebox{15pt}[15pt][10pt]}c<{$}}	
\newcolumntype{C}{>{$ \displaystyle}c<{$}} 							
\newcolumntype{L}{>{$ \displaystyle}l<{$}}  							
\newcolumntype{R}{>{$ \displaystyle}r<{$}}								

\makeatletter
\renewenvironment{thebibliography}[1]
{\section*{\refname\@mkboth{\refname}{\refname}}%
  \list{\@biblabel{\@arabic\c@enumiv}}%
       {\settowidth\labelwidth{\@biblabel{#1}}%
        \leftmargin\labelwidth
        \advance\leftmargin\labelsep
 \setlength\itemsep{1pt}			
 \setlength\baselineskip{11pt}	
        \@openbib@code
        \usecounter{enumiv}%
        \let\p@enumiv\@empty
        \renewcommand\theenumiv{\@arabic\c@enumiv}}%
  \sloppy
  \clubpenalty4000
  \@clubpenalty\clubpenalty
  \widowpenalty4000%
  \sfcode`\.\@m}
 {\def\@noitemerr
   {\@latex@warning{Empty `thebibliography' environment}}%
  \endlist}
\makeatother

\title{Phase structure of finite density QCD with a histogram method}

\ShortTitle{Phase structure of finite density QCD with a histogram method}

\author{\speaker{Y.~Nakagawa}, S.~Ejiri \\
        Graduate School of Science and Technology, Niigata University, Niigata 950-2181, Japan \\
        E-mail: \email{nakagawa@muse.sc.niigata-u.ac.jp}
        }

\author{S.~Aoki, K.~Kanaya, H.~Saito \\
        Graduate School of Pure and Applied Sciences, University of Tsukuba,
        Tsukuba, Ibaraki 305-8571, Japan
        }

\author{H.~Ohno \\
        Fakult\"{a}t f\"{u}r Physik, Universit\"{a}t,
        Bielefeld, D-33501 Bielefeld, Germany
        }

\author{T.~Hatsuda \\
        Theoretical Research Division, Nishina Center, RIKEN, Wako 351-0198, Japan
        }

\author{T.~Umeda \\
        Graduate School of Education, Hiroshima University, Hiroshima 739-8524, Japan
        }

\author{(WHOT-QCD collaboration)}

\abstract{
We study the phase structure of QCD in the $T-\mu$ plane using a histogram method
and the reweighting technique by performing phase quenched simulations of
two-flavor QCD with RG-improved gauge action and O($a$) improved Wilson quark action.
Taking the effects of the complex phase of the quark determinant using the cumulant expansion method, 
we calculate the probability distribution function of plaquette
and phase-quenched determinant as a function of $T$ and $\mu$.
We discuss the order of the QCD phase transition consulting the shape of the probability distribution function.
}

\FullConference{The 30 International Symposium on Lattice Field Theory - Lattice 2012,\\
		June 24-29, 2012\\
		Cairns, Australia}

\begin{document}

\section{Introduction}

In order to reveal the phase structure of QCD,
it is indispensable to study QCD by first principle lattice simulations.
The lattice simulations, however, have the notorious sign problem
at non-zero quark chemical potential $\mu$.
Here, we report our study on the phase structure of finite-density QCD using a histogram method with the reweighting technique \cite{Ejiri:2007ga,Saito:2011fs}
performing phase-quenched simulations and taking the effects of the complex phase of the quark determinant by the cumulant expansion. 
Results of test simulations have been reported in the previous Lattice conference \cite{Nakagawa:2011eu}. 
We found that the method is promising.
We have extended the study by increasing the number of simulation points as well as the statistics of each simulation point. 
By increasing $\mu/T$, we found a preliminary signal suggesting the on-set of the first-order transition line separating the hadronic and QGP phases.

\section{Histogram method}
\vspace{-1mm}

We define the probability distribution function $w(O)$ for an operator $\hat{O}$ by the histogram of the measurement result $O$ on the gauge configurations. 
We further define an effective potential by $V(O) = - \ln w(O)$.
Near a first-order phase transition, $w(O)$ will show a double-peak shape on finite lattices when $O$ is sensitive to the phases. 
We may thus detect first-order transition from a negative curvature between two local minima in $V(O)$.

For simplicity, let us consider the case of the degenerate $\Nf$ flavors.
Generalization to non-degenerate cases is straightforward.
To explore the phase structure at finite $T$ and $\mu$, 
we choose the generalized plaquette $\hat{P}=-\hat{S}_g/6\beta\Nsite$, 
where $\hat{S}_g$ is the gauge action and $\Nsite=N_s^3 \times N_t$ is the space-time volume, and the absolute value of the quark determinant,
$\hat{F}(\mu) = \Nf \ln \abs{\det M(\mu)/\det M(0)}$, as the operators for $w$.
We have suppressed the arguments $\kappa$ and $\beta$ in $\hat{F}$ and $M$ because we fix $\kappa$ and $c_{\rm SW}$ in this study.\footnote{
When $c_{\rm SW}$ depends on $\beta$, additional contributions to the reweighting have to be taken into account \cite{Brandt:2010uw}.
In this study, we instead fix $c_{\rm SW}$ at all $\beta$. 
This defines another renormalization scheme.
Physical properties including the phase structure will not be affected.
}

Decomposing the quark determinant as
$\RnB{\det M(\mu)}^{\Nf} = e^{i\theta(\mu)}\abs{\det M(\mu)}^{\Nf}$,
the partition function at finite $\mu$ can be written as
\begin{eqnarray}\label{eq:part_func}
\frac{Z(\beta,\mu)}{Z(\beta,0)}
= \frac{1}{Z(\beta,0)}
\int \mathscr{D}U \,e^{i\theta(\mu)} \abs{\det M(\mu)}^{\Nf} e^{6\beta\Nsite P}
&=& \int dPdF \, w_0(P,F;\beta,\mu)\,
\ave{e^{i\theta}} \nonumber \\
&=& \int dPdF\, e^{-V(P,F;\beta,\mu)},
\end{eqnarray}
where 
\begin{equation}
w_0(P,F;\beta,\mu) = \frac{1}{Z(\beta,0)}
\int \mathcal{D}U \delta(P-\hat{P})\delta(F-\hat{F}(\mu))
\abs{\det M(\mu)}^{\Nf}e^{6\beta \Nsite P}
\end{equation}
is the phase-quenched distribution function normalized by the partition function at $\mu=0$, and 
\begin{eqnarray}\label{eq:phase}
\ave{e^{i\theta}}(P,F;\mu)
&=& \frac{
\int\mathscr{D}U {e^{i\theta(\mu)}\delta(P-\hat{P})\delta(F-\hat{F}(\mu))}
\abs{\det M(\mu)}^{\Nf}e^{6\beta \Nsite P}}
{\int\mathscr{D}U \delta(P-\hat{P})\delta(F-\hat{F}(\mu))
\abs{\det M(\mu)}^{\Nf}e^{6\beta \Nsite P}} \nonumber \\
&=& \frac{
\Ave{e^{i\theta(\mu)}\delta(P-\hat{P})\delta(F-\hat{F}(\mu))}_{(\beta,\mu)}}
{\Ave{\delta(P-\hat{P})\delta(F-\hat{F}(\mu))}_{(\beta,\mu)}}
\end{eqnarray}
is the expectation value of the complex phase of the quark determinant in phase-quenched simulation 
with fixed $P$ and $F$.
The double bracket is for the expectation value in the phase-quenched simulation.
The effective potential is then given by  $V = -\ln w_0 - \ln \langle e^{i\theta} \rangle$.
Note that $\ave{e^{i\theta}}$ does not depend on $\beta$ since the factor $e^{6\beta \Nsite P}$ cancels out between numerator and denominator in \Eq{eq:phase}.

We carry out phase-quenched simulation at $(\beta_0,\mu_0)$ and calculate $w_0$ at $(\beta,\mu)$ using the reweighting technique as 
\begin{equation}\label{eq:histogram_reweighting}
w_0(P,F;\beta,\mu) = R(P,F;\beta,\beta_0,\mu,\mu_0) \; w_0(P,F;\beta_0,\mu_0),
\end{equation}
\begin{equation}\label{eq:reweighting}
R(P,F,\beta,\beta_0,\mu,\mu_0)
= e^{6(\beta-\beta_0)\Nsite P}\,\frac{
\AVE{\delta(P-\hat{P})\delta(F-\hat{F})
\abs{\frac{\det M(\mu)}{\det M(\mu_0)}}^{\Nf}}_{(\beta_0,\mu_0)}
}{
\Ave{\delta(P-\hat{P})\delta(F-\hat{F})}_{(\beta_0,\mu_0)}}.
\end{equation}
Similarly, the phase factor is given by
\begin{equation}\label{eq:reweighted_theta}
\ave{e^{i\theta}}(P,F;\mu)
= \frac{
\AVE{e^{i\theta}
\abs{\frac{\det M(\mu)}{\det M(\mu_0)}}^{\Nf}
\delta(P-\hat{P})\delta(F-\hat{F})
}_{(\beta_0,\mu_0)}
}{
\AVE{
\abs{\frac{\det M(\mu)}{\det M(\mu_0)}}^{\Nf}
\delta(P-\hat{P})\delta(F-\hat{F})
}_{(\beta_0,\mu_0)}}.
\end{equation}
where the $\beta_0$-dependence in the right hand side cancels out.
From these equations, we find that, under a shift of $\beta$ at $\mu=\mu_0$,
the slope $\partial V/\partial P$ changes by a constant factor, while $\partial^2 V/\partial P^2$ remains the same.

\section{Cumulant expansion for the complex phase of the quark determinant}
\vspace{-1mm}

Evaluation of the phase factor is suffered from the sign problem.
To mitigate the problem, we apply the cumulant expansion method to evaluate the phase factor \cite{Ejiri:2007ga,Ejiri:2009hq}:
\begin{equation}
\ave{e^{i\theta}}(P,F;\mu)
= \exp\left[ i\ave{\theta}_c
- \frac{1}{2}\ave{\theta^2}_c
- \frac{i}{3!}\ave{\theta^3}_c
+ \frac{1}{4!}\ave{\theta^4}_c
+ \cdots \right].
\end{equation}
The odd-order cumulants are the source of the sign problem.
A key observation is that, because the system has the symmetry under $\mu \leftrightarrow -\mu$, the odd-order cumulants should vanish when the statistics is sufficiently high.
Keeping the even-power cumulants only, the phase factor is real and positive.
Thus, 
the sign problem is resolved if the cumulant expansion converges.

In a previous study, we found that, when we define $\theta$ by the Taylor expansion of $ \Nf \Im \ln \det M(\mu)$ in terms of $\mu$, then the resulting distribution of $\theta$ is well described by the Gaussian function at small $\mu$ \cite{Ejiri:2007ga,Ejiri:2009hq}.
Note that $\theta$ is defined in the range $(-\infty,\infty)$, and the conventional $\theta$ modulus $2\pi$ is reproduced by taking the principal value.
With a Gaussian distribution, the cumulant expansion is dominated by the leading term $\ave{\theta^2}_c$, and thus the expansion converges.
In this study with phase-quenched simulations, we adopt the definition 
\begin{equation}\label{eq:mu_integral_theta}
\theta(\mu) 
            = \Nf \int^{\mu/T}_0 \Im \left[
              \frac{\partial(\ln\det M(\mu))}{\partial(\mu/T)}
              \right]_{\bar{\mu}} d\left( \frac{\bar{\mu}}{T} \right),
\end{equation}
{\it i.e.}, 
on each configuration generated by the phase-quenched simulation at $(\beta_0,\mu_0)$, 
we measure $\left[ \partial \ln\det M/\partial(\mu/T) \right]_{\bar\mu}$ for several values of $\bar\mu$, and calculate $\theta(\mu)$ by integrating it numerically.
For a smooth interpolation in the numerical integration procedure, we combine the information of the second derivative $\left[ \partial^2 \ln\det M/\partial(\mu/T)^2 \right]_{\bar\mu}$ too.
We confirm that the resulting distribution of $\theta$ is close to Gaussian at the simulation points we study \cite{Nakagawa:2011eu}.

Using the results of $\left[ \partial \ln\det M/\partial(\mu/T) \right]_{\bar\mu}$ and $\left[ \partial^2 \ln\det M/\partial(\mu/T)^2 \right]_{\bar\mu}$, we also calculate $F(\mu)$ and the reweighting factor $\left| \det M(\mu)/\det M(\mu_0) \right|$ in \Eqs{eq:reweighting} and (\ref{eq:reweighted_theta}) as follows:
\begin{eqnarray}
           \label{eq:mu_integral_F}
F(\mu) &=& \Nf \ln \abs{\frac{\det M(\mu)}{\det M(0)}}
        =  \Nf \int^{\mu/T}_0 \Re \left[
           \frac{\partial(\ln\det M(\mu))}{\partial(\mu/T)}
           \right]_{\bar{\mu}} d\left( \frac{\bar{\mu}}{T} \right), \\
           \label{eq:mu_integral_C}
C(\mu) &=& \Nf \ln \abs{\frac{\det M(\mu)}{\det M(\mu_0)}}
        =  \Nf \int^{\mu/T}_{\mu_0/T} \Re \left[
           \frac{\partial(\ln\det M(\mu))}{\partial(\mu/T)}
           \right]_{\bar{\mu}} d\left( \frac{\bar{\mu}}{T} \right).
\end{eqnarray}


To the lowest order of the cumulant expansion,
the curvatures of the effective potential $V$ in $P$ and $F$ directions are now given by
\begin{eqnarray}
\frac{\partial^2 V}{\partial P^2}(P,F;\beta,\mu)
&=& \frac{\partial^2 (-\ln w_0)}{\partial P^2}(P,F;\beta_0,\mu_0)
- \frac{\partial^2 \ln R}{\partial P^2}(P,F;\beta,\beta_0,\mu,\mu_0)
+ \frac{1}{2}\frac{\partial^2 \ave{\theta^2}_c}{\partial P^2}(P,F;\mu), \nonumber\\
\frac{\partial^2 V}{\partial F^2}(P,F;\beta,\mu)
&=& \frac{\partial^2 (-\ln w_0)}{\partial F^2}(P,F;\beta_0,\mu_0)
- \frac{\partial^2 \ln R}{\partial F^2}(P,F;\beta,\beta_0,\mu,\mu_0)
+ \frac{1}{2}\frac{\partial^2 \ave{\theta^2}_c}{\partial F^2}(P,F;\mu). \nonumber \\
\label{eq:curvature}
\end{eqnarray}
If the curvature of $V$ is always positive, the transition between confined and deconfined phases is not first order.
On the other hand, if negative curvature appears in the critical region, it indicates on-set of first-order transition.

\begin{figure}[btp]
\begin{center}
\begin{minipage}{0.46\hsize}
\resizedgraphics{1.0}{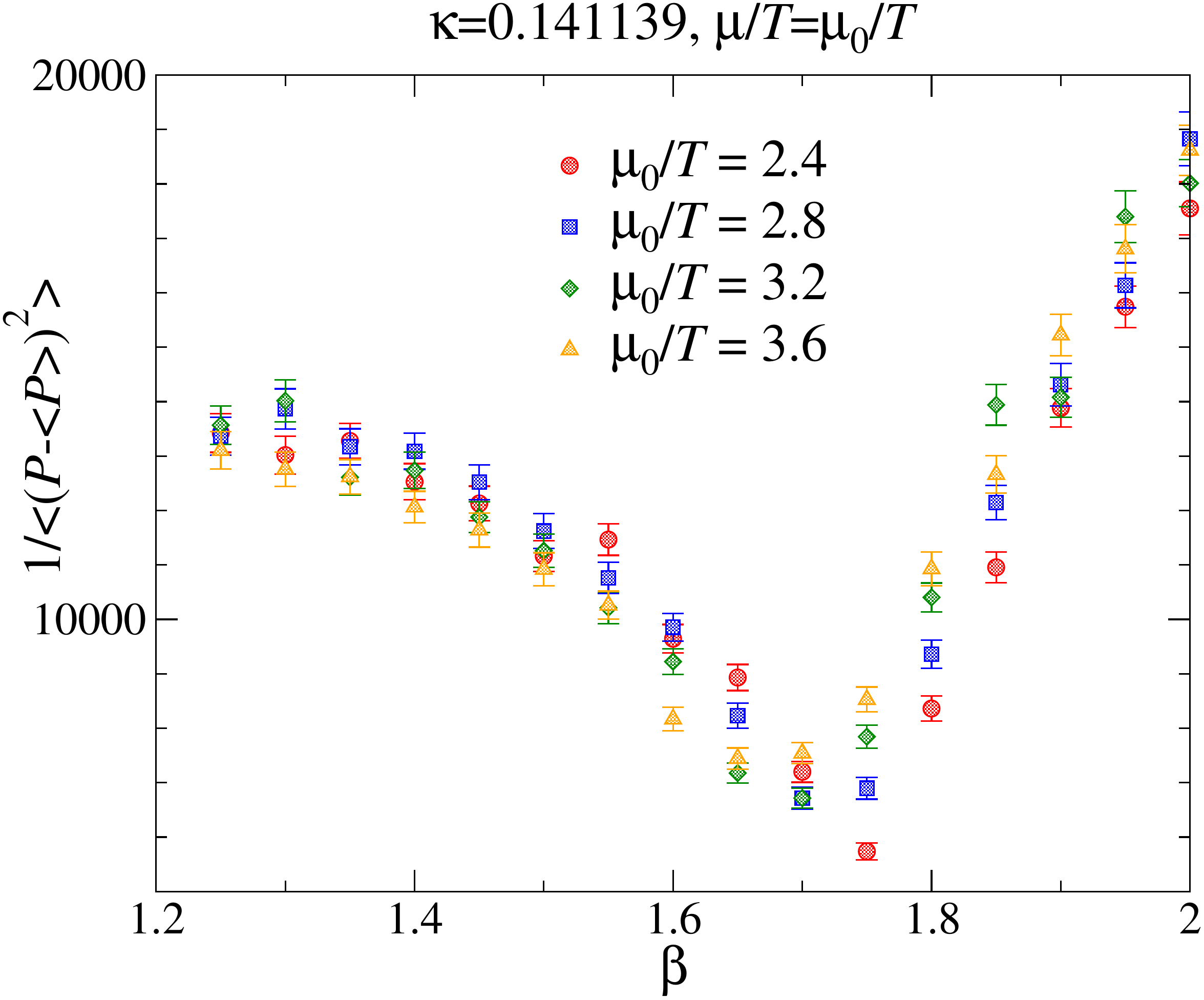}
\end{minipage}
\hspace{0.02\hsize}
\begin{minipage}{0.46\hsize}
\resizedgraphics{1.0}{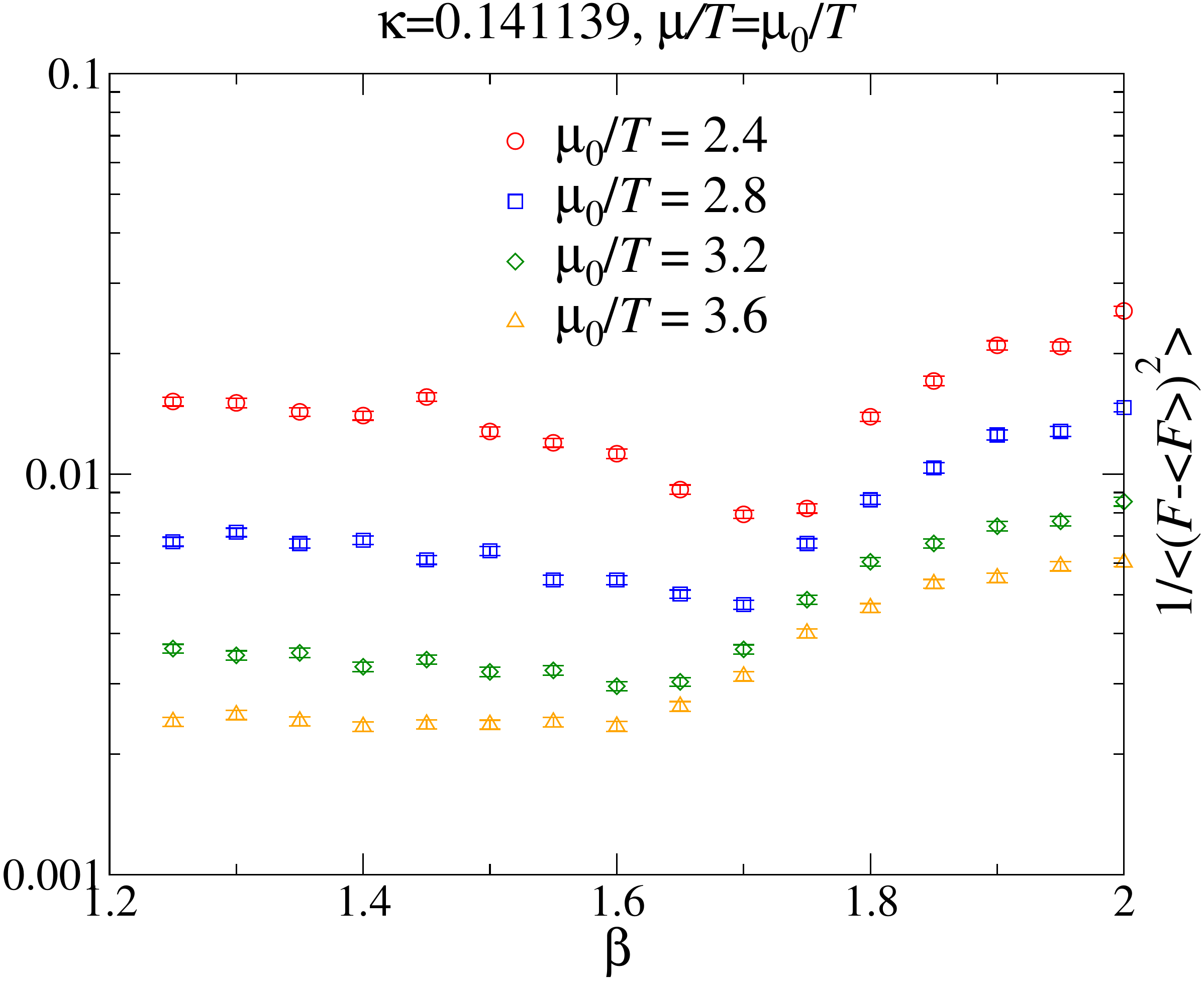}
\end{minipage}
\end{center}
\vspace{-8mm}
\caption{
The inverse susceptibilities of $P$ (left panel) and $F$ (right panel)
as a function of $\beta$ at various $\mu_0/T$.
}
\label{fig:inverse_suceptibilities}
\end{figure}

\section{Numerical simulations and the results}
\vspace{-1mm}

We use the RG-improved Iwasaki gauge action and the $O(a)$-improved Wilson quark action.
We generated gauge configurations on $8^3\times 4$ lattice
 in  two-flavor QCD with the phase-quenched measure at various values of $(\beta_0,\mu_0)$ in the range $\beta_0=1.2$--2.0 and $\mu_0/T=2.0$--4.0.
$\kappa$ and $c_{SW}$ are fixed to 0.141139 and 1.603830 respectively.
We measure the first and the second derivatives of $\ln \det M$ every 10 trajectories
using the random noise method of \cite{Ejiri:2009hq} with 50 noises.
The statistics is $2900\times10$ trajectories after thermalization.

\begin{figure}[hbtp]
\begin{center}
\begin{minipage}{0.46\hsize}
\resizedgraphics{1.0}{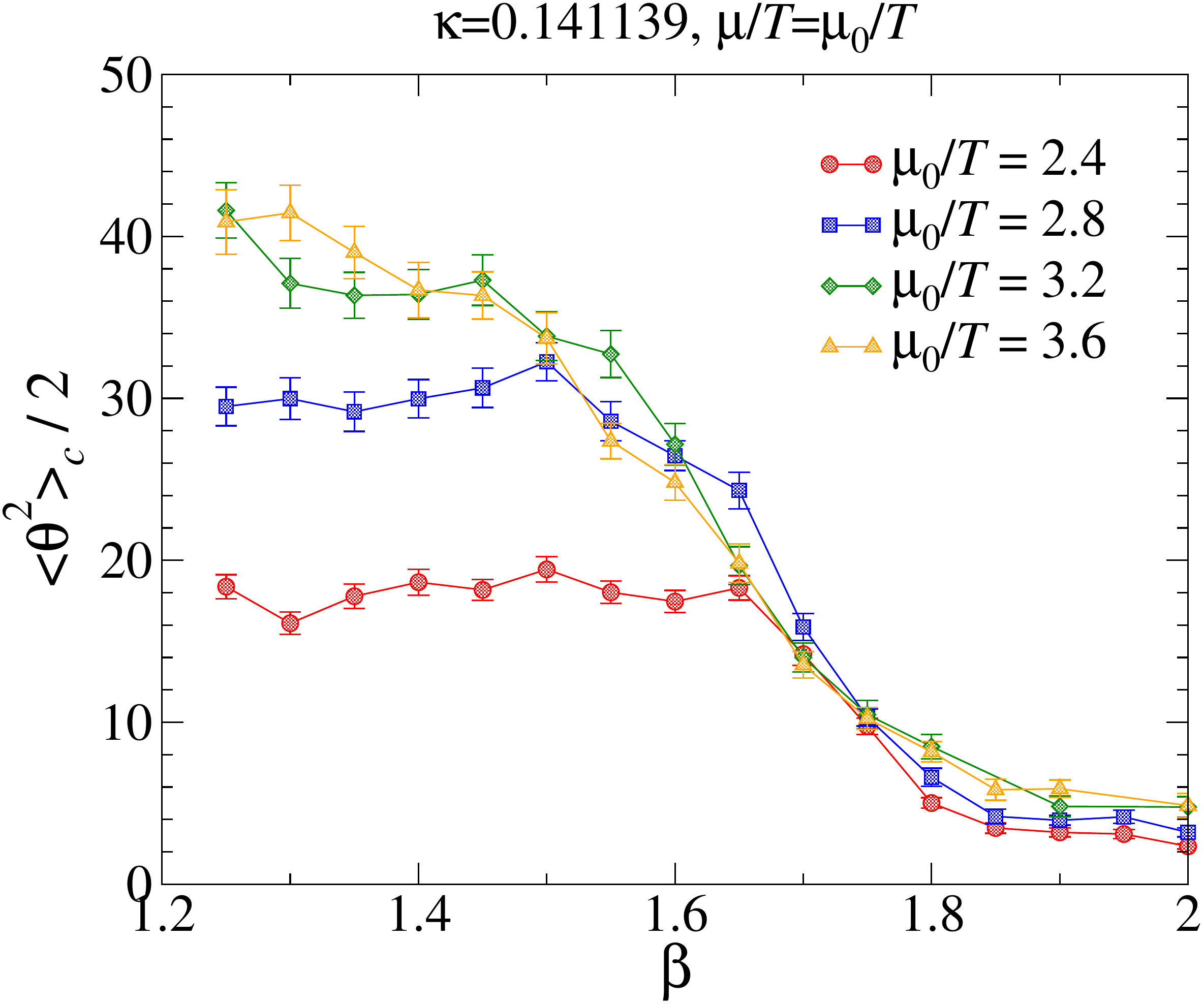}
\end{minipage}
\hspace{0.01\hsize}
\begin{minipage}{0.46\hsize}
\resizedgraphics{1.05}{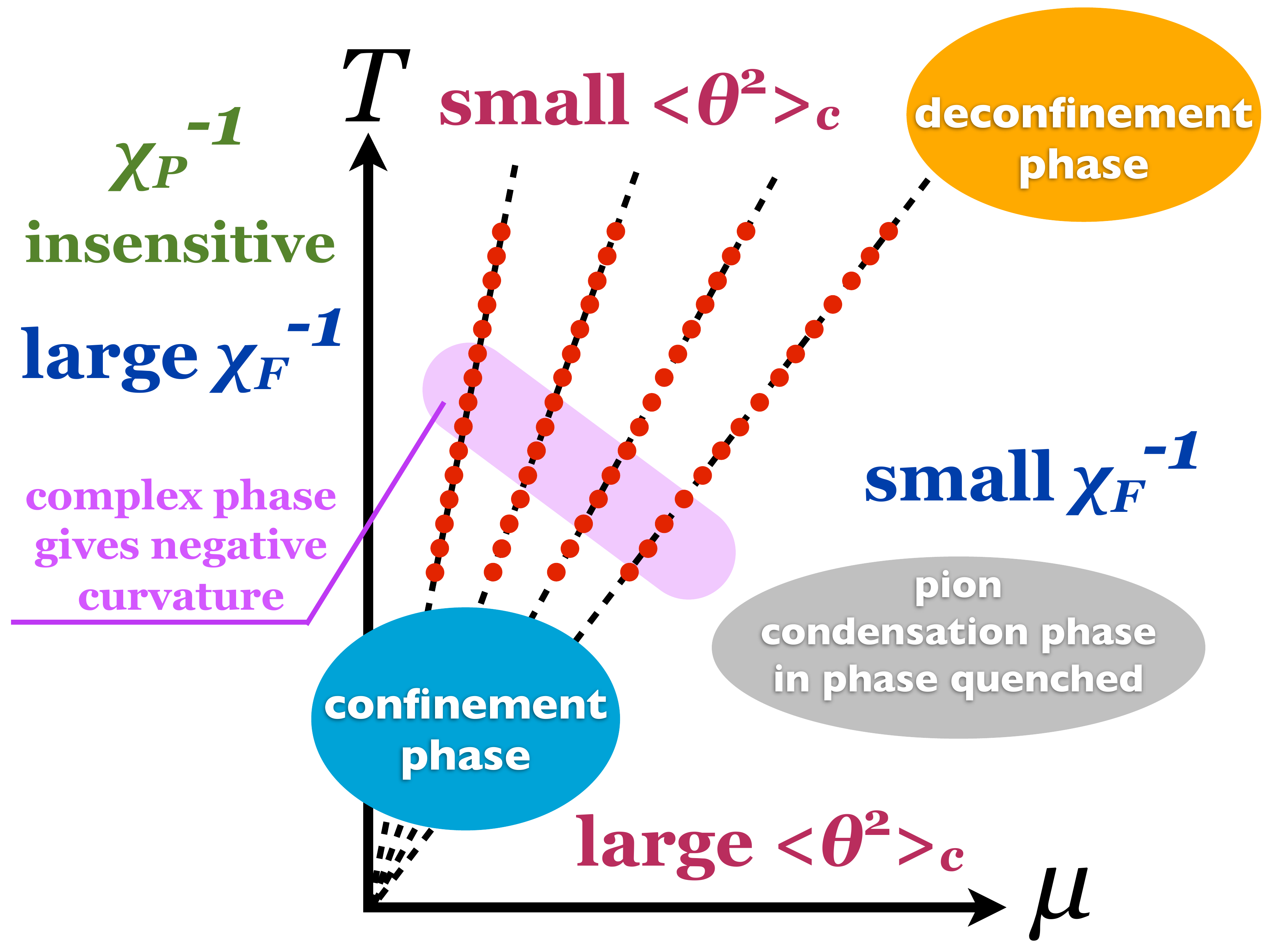}
\end{minipage}
\end{center}
\vspace{-8mm}
\caption{
Leftl:
the statistical average of the second order cumulant of the complex phase
as a function of $\beta$.
Right: 
Schematic summary of the situation in the $T-\mu$ plane.
Red dots represent the simulation points.
}
\label{fig:cumulants}
\end{figure}

We find that the phase-quenched distributions for $P$ and $F$ around $\ave{P}_{(\beta_0,\mu_0)}$
and $\ave{F}_{(\beta_0,\mu_0)}$ have simple single bump forms. 
By approximating them by Gaussian functions, we estimate the second derivatives of $\ln w_0$ 
in \Eq{eq:curvature} by inverse susceptibilities of $P$ and $F$ as
\begin{equation}
\frac{\partial^2 (-\ln w_0)}{\partial P^2} = \frac{1}{\ave{(P-\ave{P})^2}} = \frac{1}{\chi_P}
, \hspace{5mm}
\frac{\partial^2 (-\ln w_0)}{\partial F^2} = \frac{1}{\ave{(F-\ave{F})^2}} = \frac{1}{\chi_F},
\end{equation}
where $\ave{\;\cdot\;}$ is the phase-quenched average at $(\beta_0,\mu_0)$.

The inverse susceptibilities are shown in \Fig{fig:inverse_suceptibilities}
as function $\beta$ at various $\mu_0$.
We find that $\chi_F$ increases rapidly as $\mu_0$ increases. 
At small $\beta$, $\ave{(F-\ave{F})^2}$ at $\mu_0/T=3.6$ is about 10 times smaller than that at $\mu_0/T=2.4$.
A model calculation with an isotriplet chemical potential, which corresponds to the phase quenched two-flavor QCD,
suggests that there is a second-order phase transition
to the pion condensed phase at large $\mu_0$ \cite{Son:2000xc}.
The increase of $\chi_F$ suggests that the system is getting closer to the pion condensed phase as $\mu_0$ is increased.
On the other hand, $\chi_P$ is insensitive to $\mu_0$.
It is hard to expect a region with negative curvature in the $P$ direction.
We thus concentrate on the curvature in the $F$ direction in the followings.

\begin{figure}[btp]
   \begin{minipage}{0.33\hsize}\begin{center}
   \resizedgraphics{1.0}{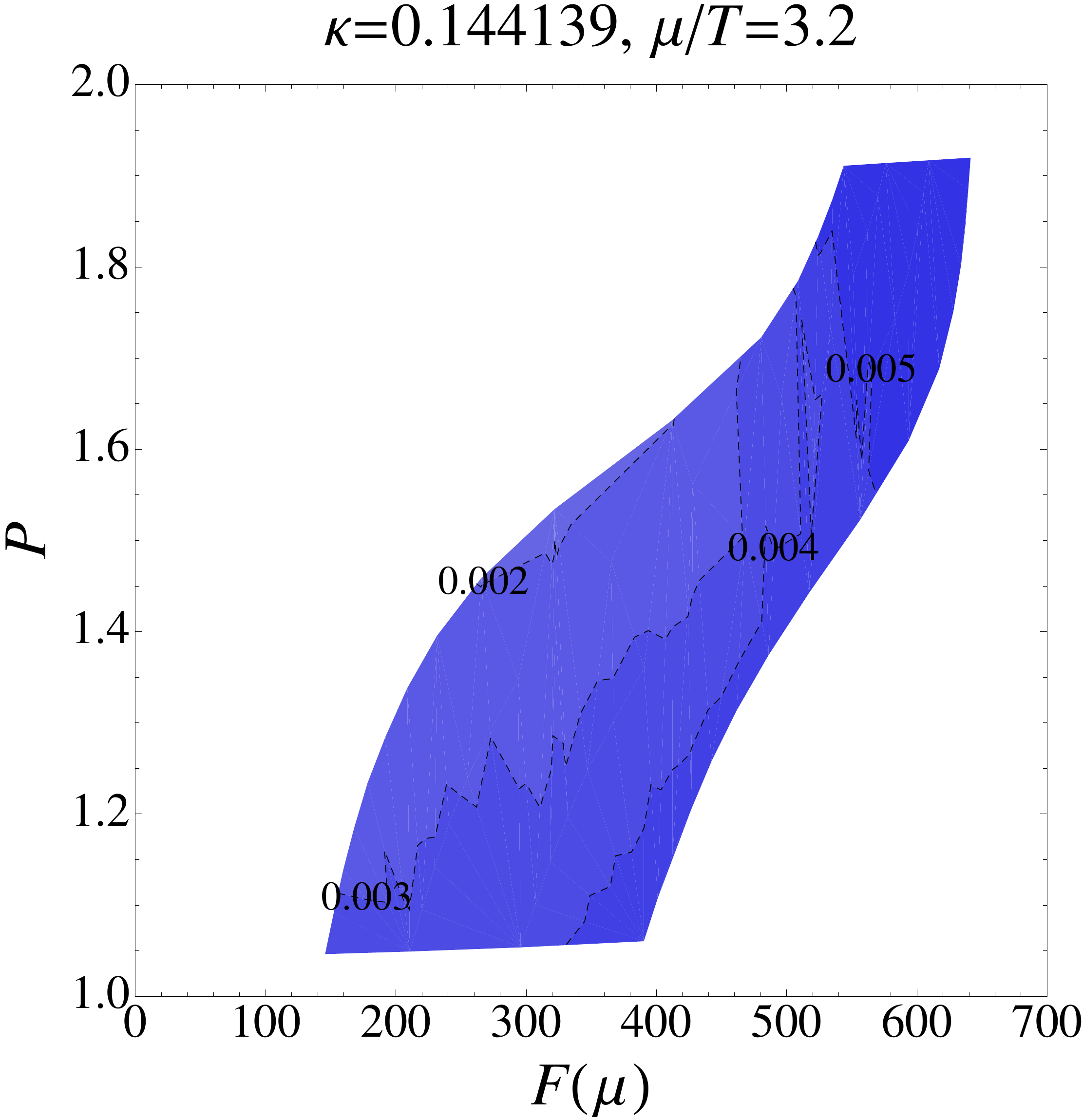}
   \end{center}\end{minipage}
   \begin{minipage}{0.33\hsize}\begin{center}
   \resizedgraphics{1.0}{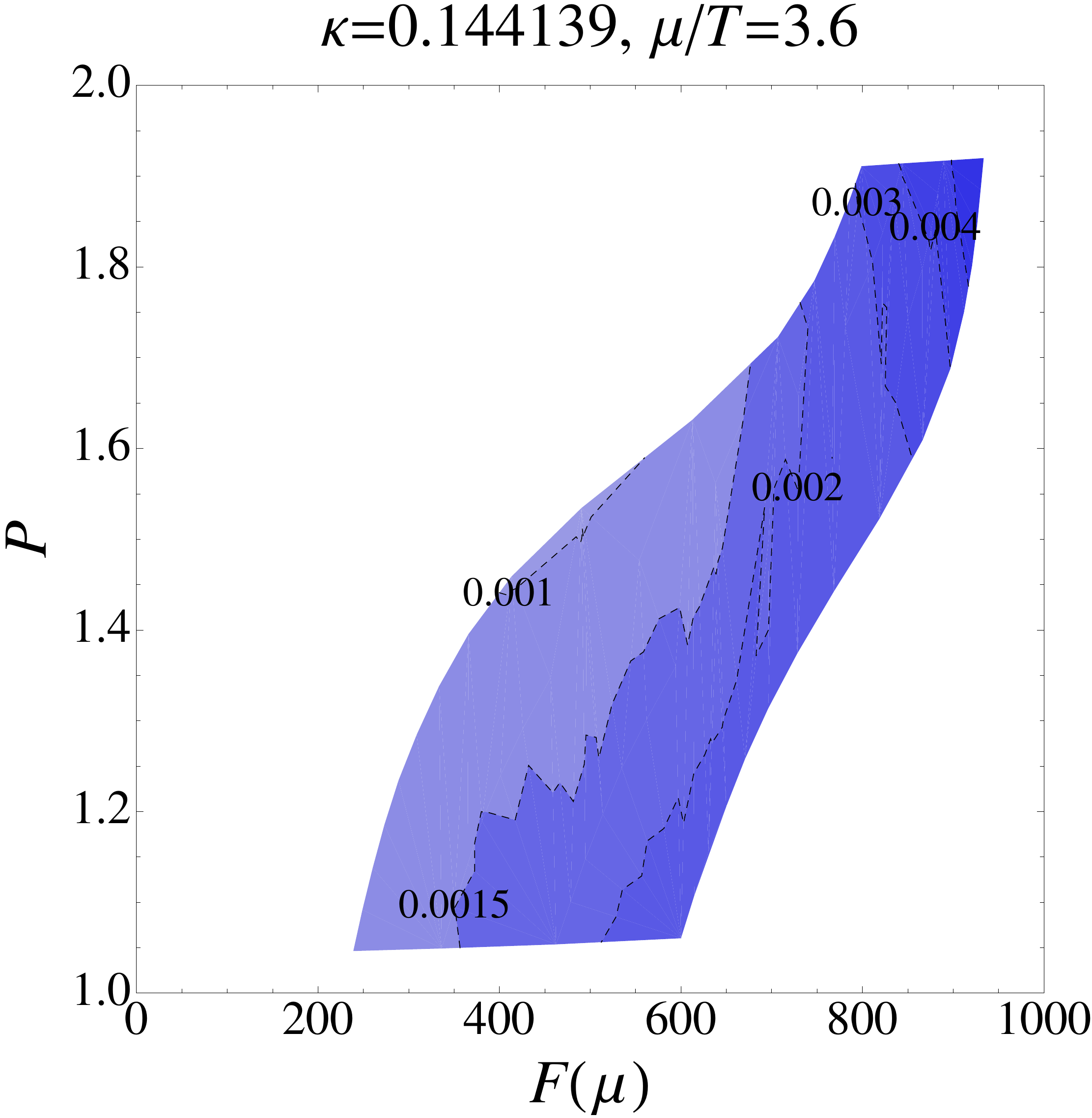}
   \end{center}\end{minipage}
   \begin{minipage}{0.33\hsize}\begin{center}
   \resizedgraphics{1.0}{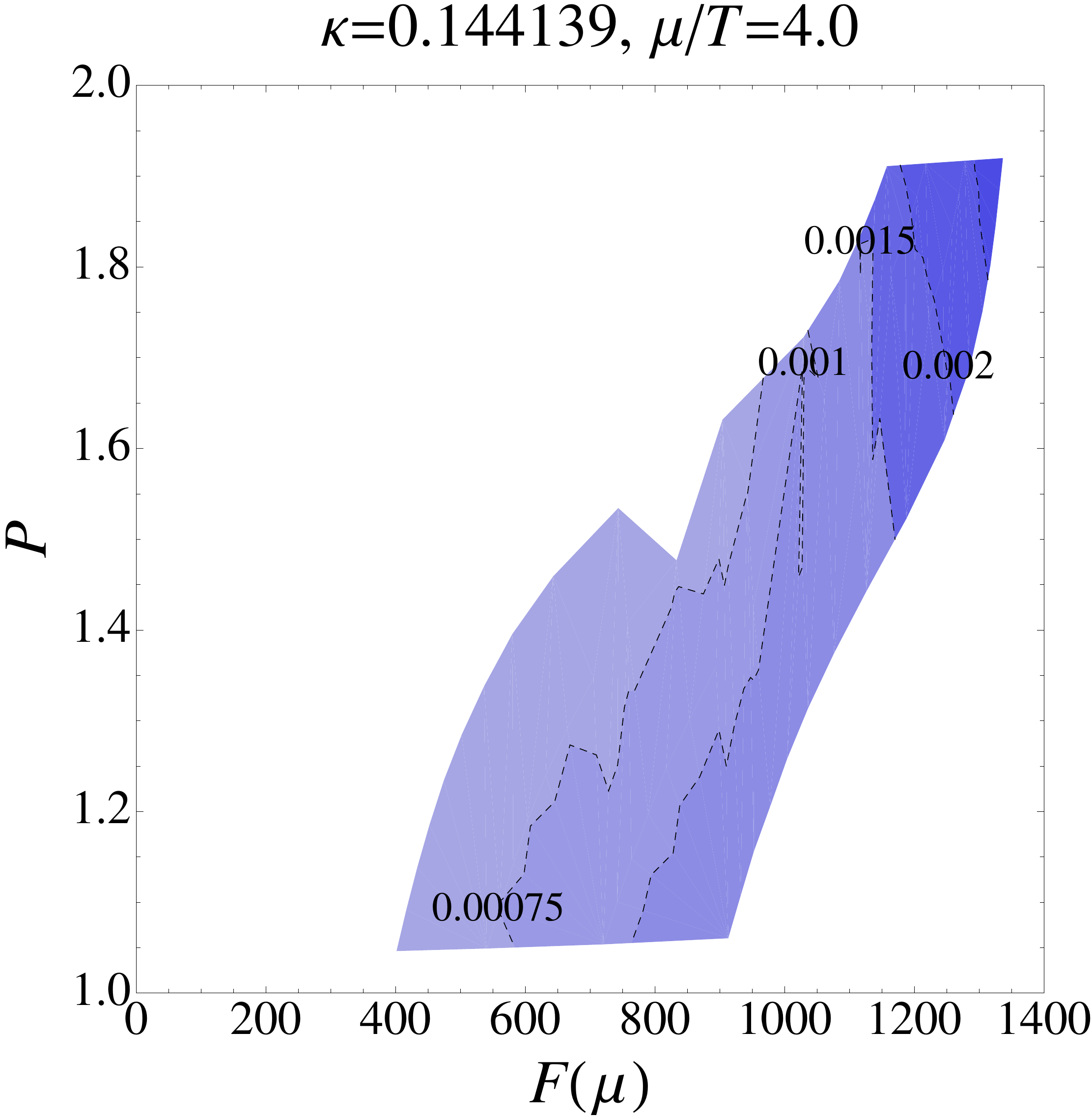}
   \end{center}\end{minipage}
   \caption{
   Contour plot of the curvature of the histogram, $\partial^2(-\ln w_0)/\partial F^2$,
   at $\mu/T=3.2, 3.6, 4.0$.
   }
   \label{fig:Contour_d2V0dF2}
   \vspace{4mm}
%
   \begin{minipage}{0.33\hsize}\begin{center}
   \resizedgraphics{1.0}{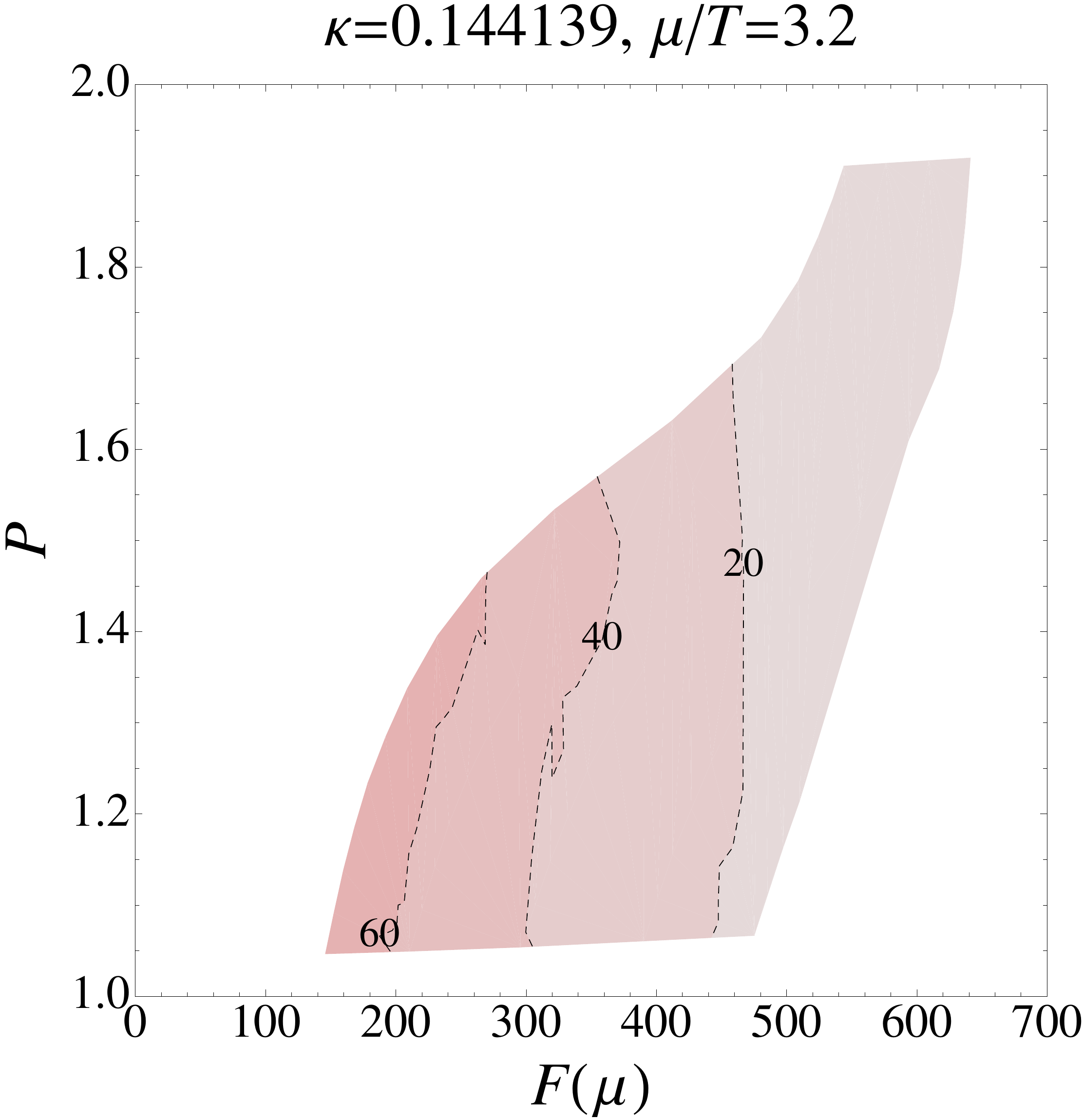}
   \end{center}\end{minipage}
   \begin{minipage}{0.33\hsize}\begin{center}
   \resizedgraphics{1.0}{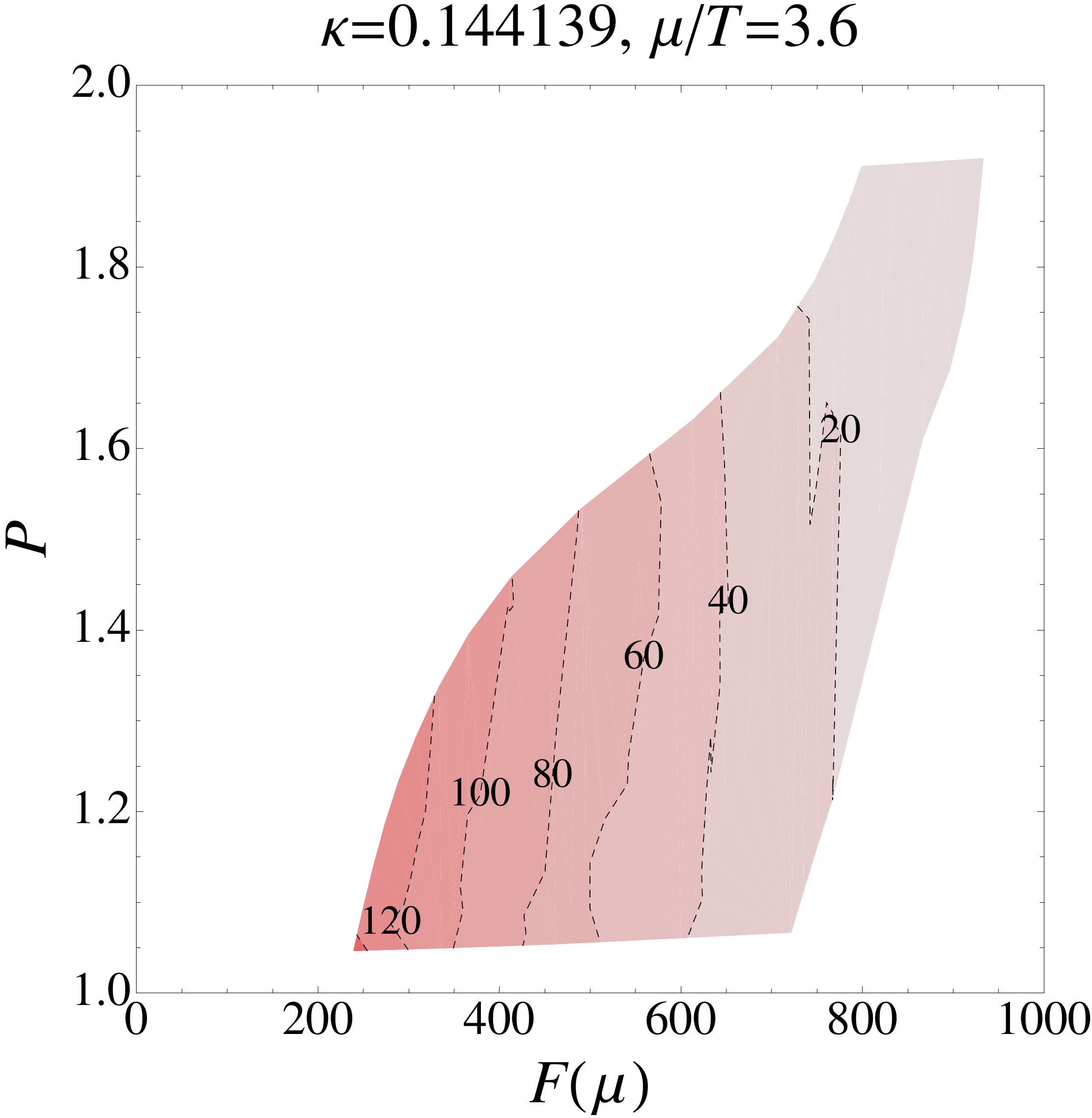}
   \end{center}\end{minipage}
   \begin{minipage}{0.33\hsize}\begin{center}
   \resizedgraphics{1.0}{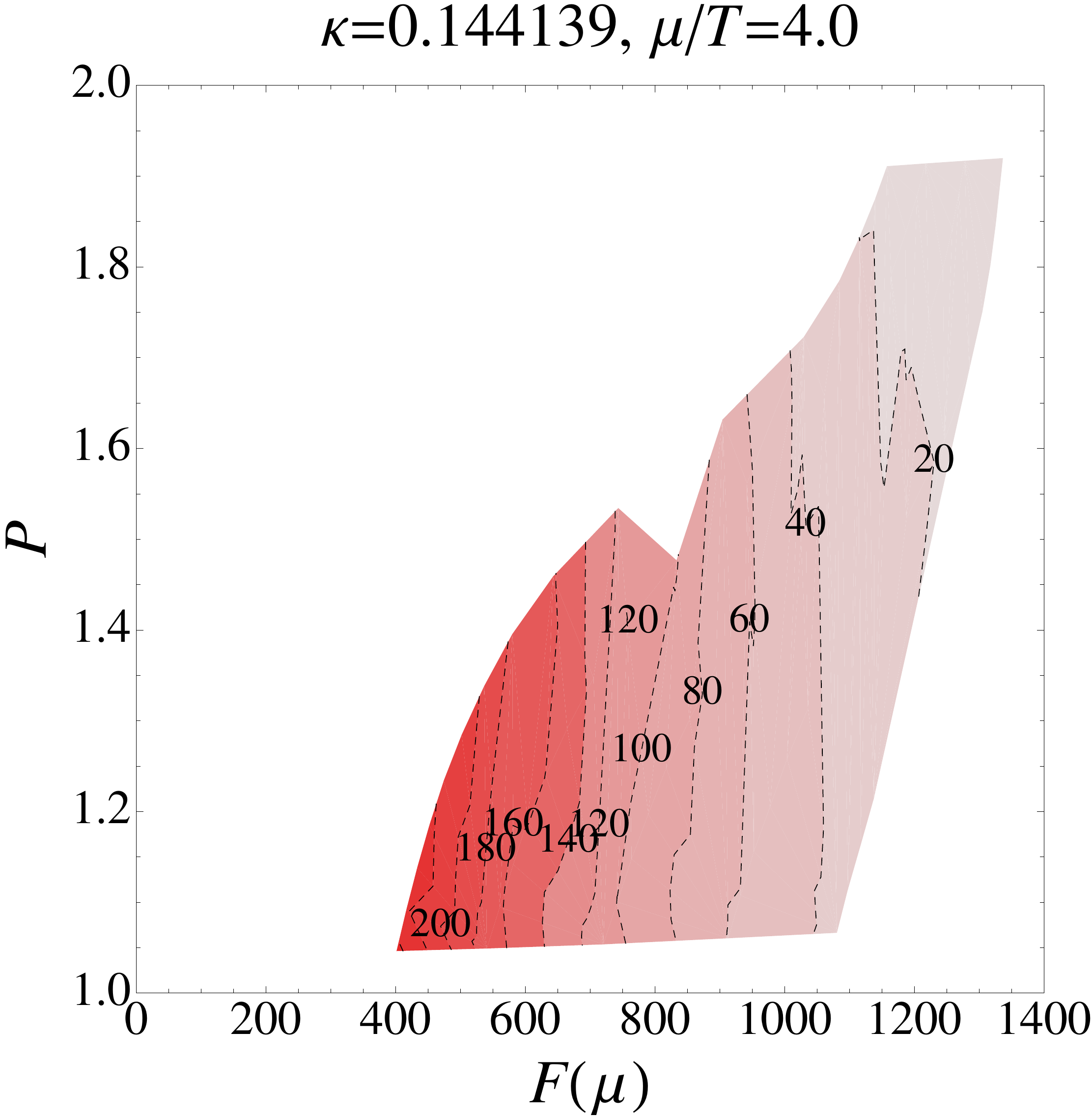}
   \end{center}\end{minipage}
   \caption{
   Contour plot of the second order cumulant $\frac{1}{2}\ave{\theta^2}_c$
   at $\mu/T=3.2, 3.6, 4.0$.
   }
   \label{fig:Contour_ThtScnd}
\end{figure}

\Fig{fig:Contour_d2V0dF2} shows the results of $\partial^2(-\ln w_0)/\partial F^2$ evaluated at various $(\beta_0,\mu_0)$
in the $P$--$F$ plane for $\mu/T=3.2$, 3.6, and 4.0.
The distribution function gives positive contribution and decreases quickly as the chemical potential increases, in the range we have studied.

To evaluate $\ln R$ and $\ave{\theta^2}_c$ with \Eqs{eq:reweighting} and (\ref{eq:reweighted_theta}), we calculate the reweighting factor $ |\det M(\mu)/\det M(\mu_0)|^{\Nf} = e^{C}$. 
We note that, in the range of parameters we have studied except just at $\mu=\mu_0$ where the reweighting factor is not needed, $C$ shows a strong linear correlation with $F$ such that $C \approx aF+b$ is well satisfied in the ensemble at $(\beta_0,\mu_0)$.
When we approximate $C \approx aF+b$, the reweighting factor $e^C$ can be factored out from the ensemble average with constrained $F$. 
Then, the resulting $\ln R$ is a linear function of $F$, and $\ln R$ does not contribute to $\partial^2 V/\partial F^2$.

With this approximation, we can also evaluate $\ave{\theta^2}_c$ without the reweighting factor. 
In this study, we roughly estimate $\ave{\theta^2}_c$ by the statistical average at the simulation point. 
In the study of curvatures, this corresponds to an approximation that $\theta$ depends on $P$ and $F$ at most linearly in the dominant range around $\ave{P}$ and $\ave{F}$ at each simulation point.
The results of $\ave{\theta^2}_c$ are plotted in the left panel of \Fig{fig:cumulants} as a function of $\beta$.
We observe that $\ave{\theta^2}_c$ 
shows a sharp bend downwards near the pseudo critical line when $\mu_0$ is large.
When we translate the $\beta$-dependence to $F$ and $P$ dependences, this means that $\ave{\theta^2}_c$ gives a negative contribution to the curvature of $V$.
The second order cumulant in $P-F$ plane is drawn in \Fig{fig:Contour_ThtScnd}.
We find that the fluctuation of the phase of the quark determinant increases
in the small $(P,F)$ region, and the contours get closer to each other
at larger $\mu/T$.
When the negative contribution becomes sufficiently strong, we may expect a negative curvature of $V$.
This situation is depicted schematically in the right panel
of \Fig{fig:cumulants}.

%
\begin{figure}[btp]
   \begin{minipage}{0.33\hsize}\begin{center}
   \resizedgraphics{1.0}{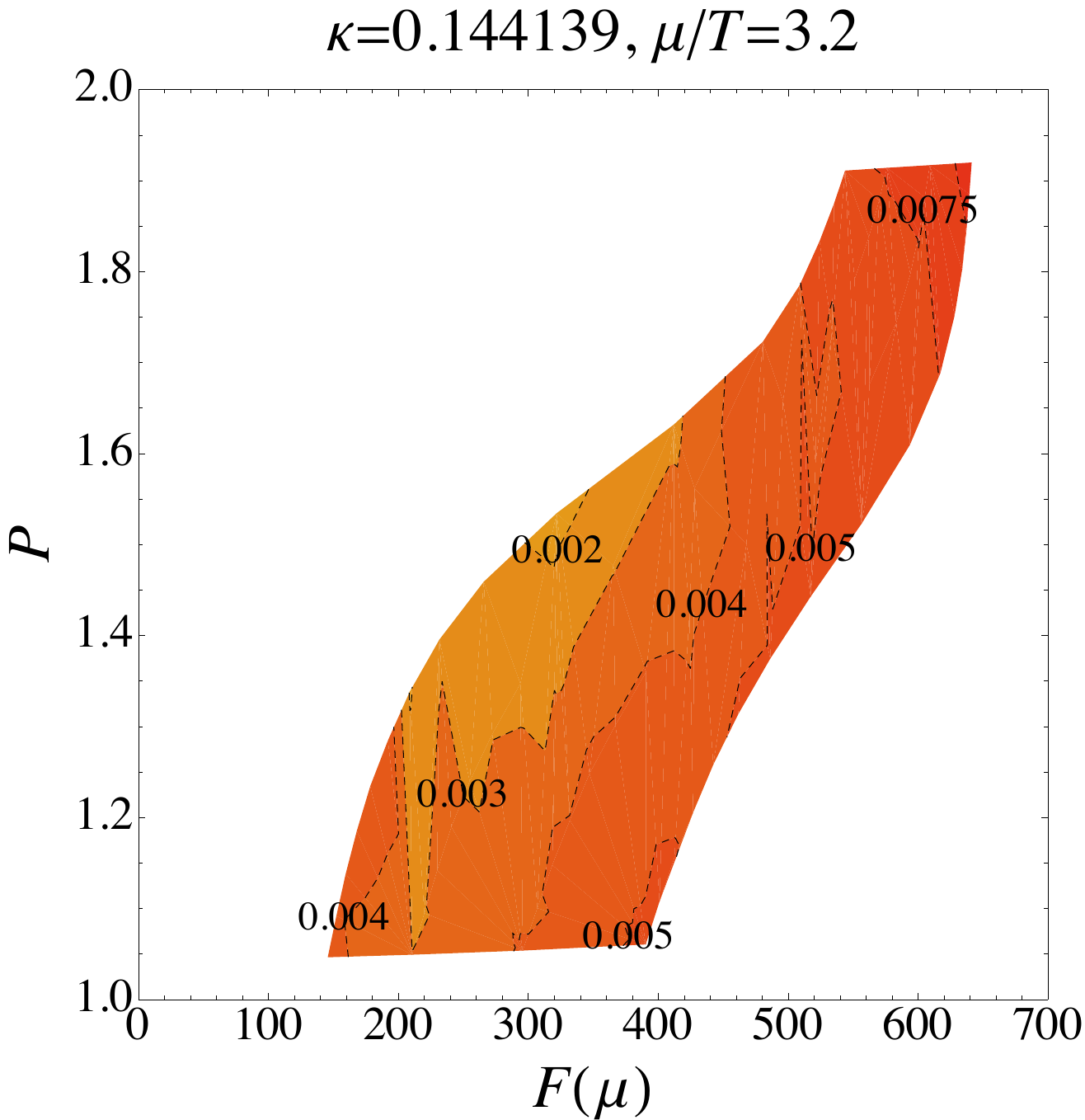}
   \end{center}\end{minipage}
   \begin{minipage}{0.33\hsize}\begin{center}
   \resizedgraphics{1.0}{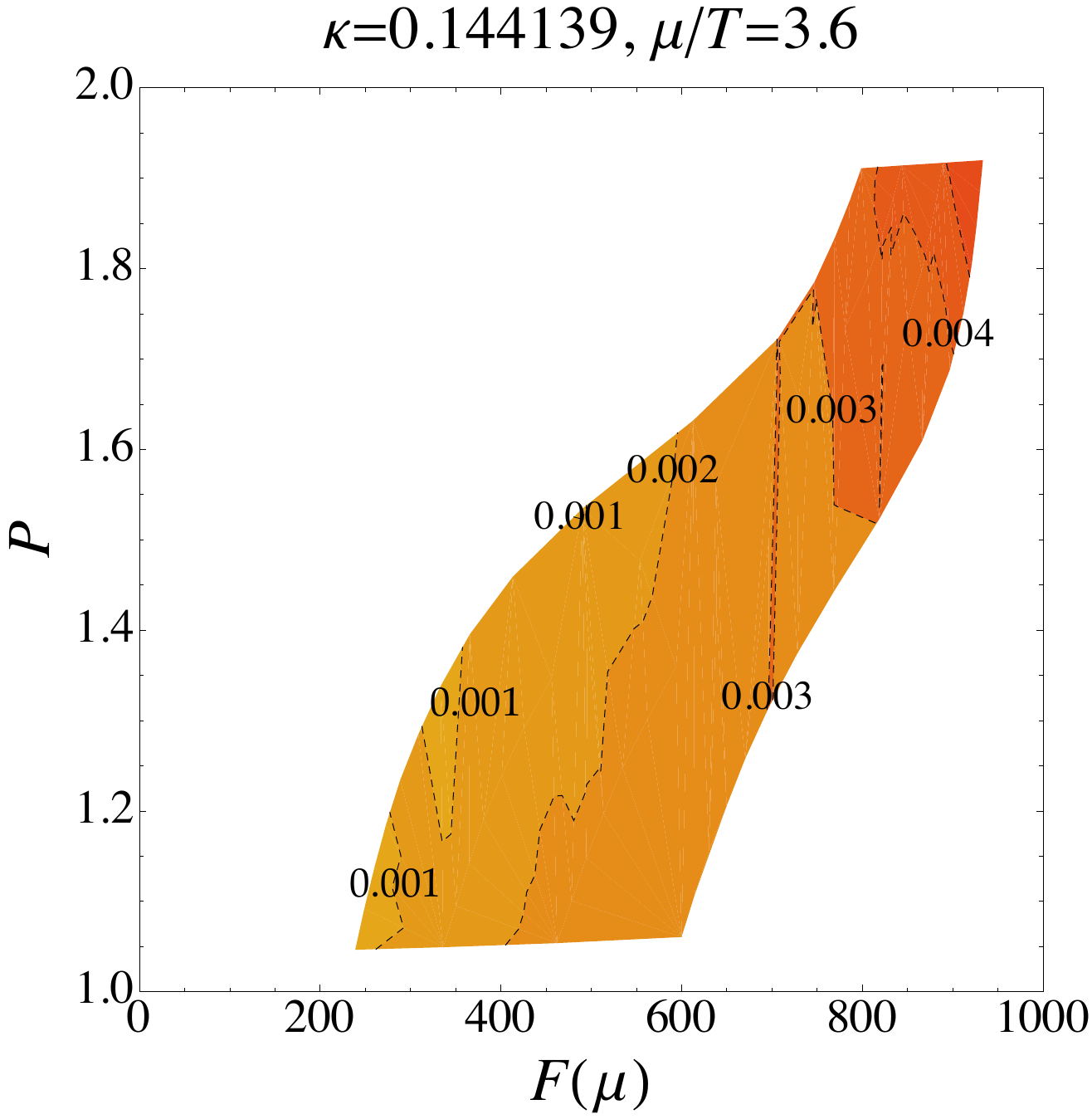}
   \end{center}\end{minipage}
   \begin{minipage}{0.33\hsize}\begin{center}
   \resizedgraphics{1.0}{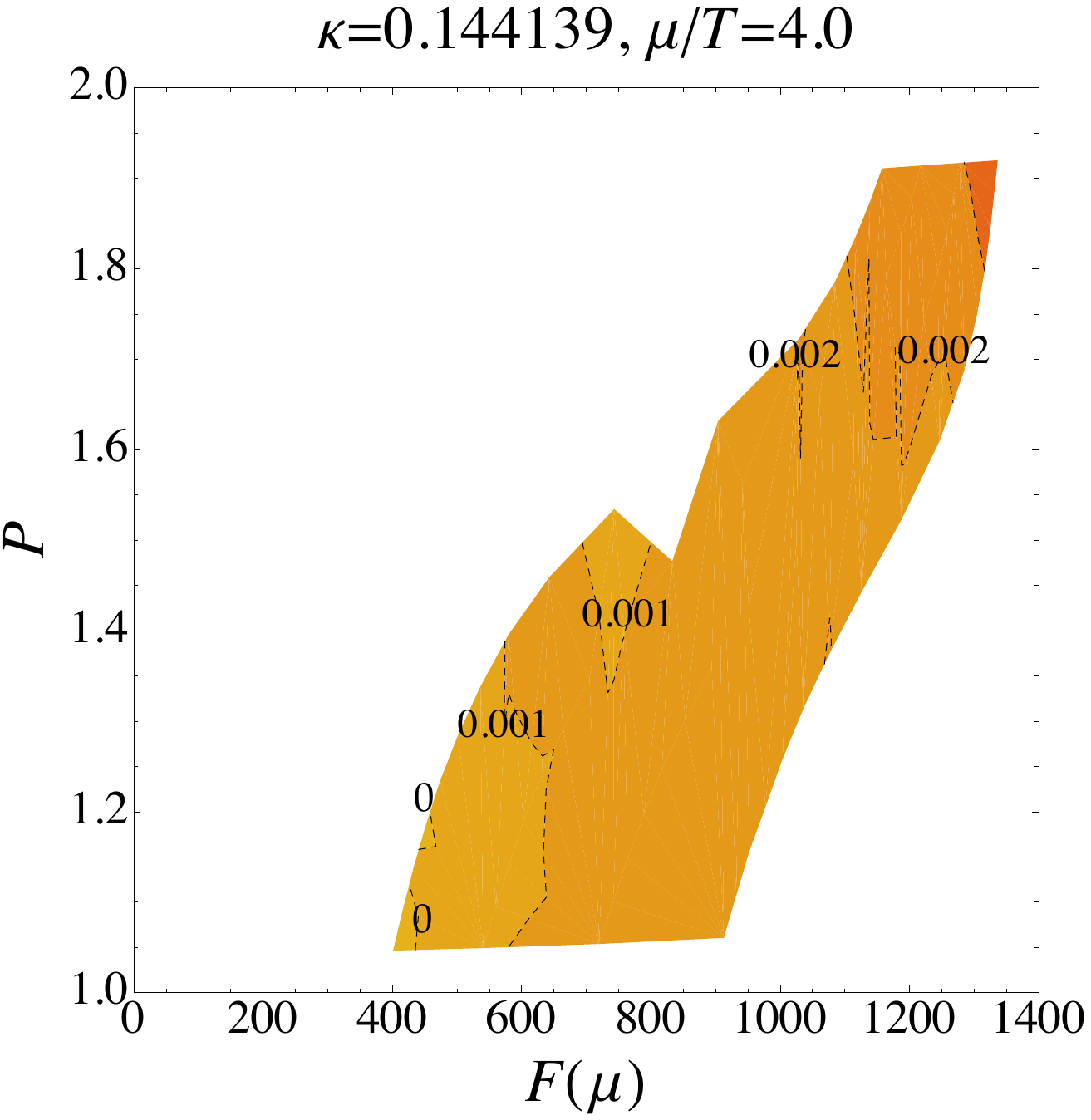}
   \end{center}\end{minipage}
   \caption{
   Contour plot of the curvature of the effective potential in $F$ direction
   at $\mu/T=3.2, 3.6, 4.0$.
   }
   \label{fig:Contour_dVeffdF2}
\end{figure}

\Fig{fig:Contour_dVeffdF2} shows the final results for the curvature of the effective potential
in the $F$ direction at $\mu/T=3.2, 3.6, 4.0$.
We observe that the curvature remains positive in $P-F$ plane at $\mu/T=3.2$ and $3.6$.
At $\mu/T=4.0$, however, we see a region at small $(P,F)$ in which the curvature is consistent with zero.
Although we could not find a region with negative curvature, the appearance of zero curvature region suggests that we have the end point of the first-order transition line separating the hadronic and QGP phases around this point.

\section{Summary}
\vspace{-1mm}

We proposed a new approach to explore the phase structure of finite density QCD
based on the histogram method and the reweighting technique
using phase quenched simulations.
We defined an effective potential in terms of the probability distribution function for a generalized plaquette (corresponding to the gauge action) and the absolute value of the quark determinant (corresponding to the quark action), and studied the curvature of the effective potential to detect the first order phase transition.
We carried out a phase-quenched simulation of two-flavor QCD with improved Wilson quarks, and incorporated the effects of the complex phase of the quark determinant by the reweighting technique and the cumulant expansion method.
We found that the complex phase factor contributes negatively to the curvature of the effective potential.
If the curvature of the effective potential becomes negative around the minimum of the potential, the transition is first order.
Although we could not find a region with negative curvature within the present range of simulation parameters, when we increase $\mu/T$ up to 4.0, we observe a region in which the curvature is consistent with zero.
This suggests the end point of the first-order transition line separating the hadronic and QGP phases around this point.

\vspace{3mm}

This work is in part supported by Grants-in-Aid of the Japanese Ministry of Education,
Culture, Sports, Science and Technology,
(Nos.20340047, 21340049, 22740168, 23540295)
and by the Grant-in-Aid for Scientific Research on Innovative Areas
(Nos. 20105001, 20105003, 23105706).

\end{document}